\documentclass[a4paper,11pt]{article} 
\usepackage{jheppub}

\usepackage[utf8]{inputenc}
\usepackage{graphicx}
\usepackage{csquotes}
\usepackage{xcolor}
\usepackage{amsmath}
\usepackage{amssymb}
\usepackage{mathtools}
\usepackage{placeins}
\usepackage{derivative}
\usepackage{bm}
\usepackage{braket}
\usepackage{physics}
\usepackage[low-sup]{subdepth}
\usepackage{enumitem}
\usepackage{cleveref}
\crefname{subsection}{subsection}{subsections}
\Crefname{subsection}{Subsection}{Subsections}
\usepackage{suffix} 
\usepackage{booktabs}
\usepackage{tabu} 
\usepackage{diagbox}

\newcommand{\Eqref}[1]{Eq.~\eqref{#1}}
\newcommand*\diff{\mathop{}\!\mathrm{d}}
\newcommand{\rangleE}{\rangle_{_{\!\text E}}}
\newcommand{\rangleBeta}{\rangle_{_{\!\beta}}}
\newcommand{\tflow}{t_{\text f}}
\newcommand{\Gtf}{G_{t_{\text f}}}
\newcommand{\chitop}{\chi_{\text{top}}}
\newcommand{\Vlatt}{V_{\ell}}
\newcommand{\Gret}{G_{\text{R}}}
\newcommand{\GEucl}{G_{\text{E}}}
\newcommand{\Hret}{H_{\text{R}}}
\newcommand{\HE}{H_{\text{E}}}
\newcommand{\HEtf}{H_{\text{E},\tflow}}
\newcommand{\hEtf}{h_{\text{E},\tflow}}
\newcommand{\alphas}{\alpha_{\text{s}}}

\begin{document}

\title{Thermal axion production in QCD from the lattice}

\author{Luis Neubauer, Guy D. Moore}
\affiliation{Institut f\"ur Kernphysik, Technische Universit\"at Darmstadt\\
Schlossgartenstra{\ss}e 2, D-64289 Darmstadt, Germany}
\emailAdd{luis.neubauer@stud.tu-darmstadt.de,guy\_david.moore@tu-darmstadt.de}

\date{\today}

 \abstract{Thermal axion production in the early Universe is controlled by the QCD topological-charge-density spectral function evaluated on the light-cone as a function of momentum $k$.
 We show how weighted integrals of this quantity can be computed using Euclidean lattice QCD without any analytic continuation, similar to what Harvey Meyer and collaborators have recently done for the photon production rate \cite{meyer-photon-2018,meyer-photon-2024}.
 We then test the approach in pure-glue QCD at $T=1.5 T_c$, finding that it is feasible to extract at least one and possibly two finite weighted $k$-integrals of the $k$-dependent production rate.
 This opens the possibility for extractions in full QCD at temperatures around and somewhat above the crossover temperature, where other methods are not reliable.}


\maketitle

\section{Introduction}
\label{sec:intro}

The QCD axion
\cite{Peccei:1977hh,Peccei:1977ur,Weinberg:1977ma,Wilczek:1977pj}
is a well-motivated hypothetical particle which could explain why the theory of the strong interactions, QCD, does not feature large violations of time-reversal invariance
\cite{tHooft:1976,Jackiw:1976pf,Callan:1979bg}.
While the couplings of a QCD axion are in general model dependent, one coupling is not; the essential feature of the QCD axion is that it couples to the topological charge density.
Writing the topological charge density of QCD as%
\footnote{The gauge group is SU(3) and its gauge fields are $G_\mu = G_\mu^A T^A$ with $T^A$ Hermitian and in the fundamental-representation with the normalization convention $\text{Tr}\: T^A T^B = \delta^{AB}/2$.
Gauge fields are in the geometrical normalization in which the covariant derivative is 
$D_\mu = \partial_\mu -i G_\mu$; the gauge coupling appears in the denominator of the field-strength term in the action.
The field strength is $G_{\mu\nu} = -i [D_\mu\,,\, D_\nu]$.
The metric is mostly-positive.}
\begin{align}
\label{qdef}
    q(x) & \equiv \frac{1}{32\pi^2} \epsilon^{\mu\nu\alpha\beta} \, \text{Tr}\, G_{\mu\nu} G_{\alpha\beta},
\end{align}
the axion field couples to QCD through the interaction Lagrangian
\begin{align}
\label{axionLagrangian}
    -\mathcal{L}_a & = \frac 12 \partial_\mu a\, \partial^\mu a + \frac{a}{f_a} q + \ldots 
\end{align}
where $\ldots$ represents model-dependent additional couplings to electromagnetism and to fermions.

\begin{figure}[bth]
    \centerline{\includegraphics[width=0.5\textwidth,angle=270]{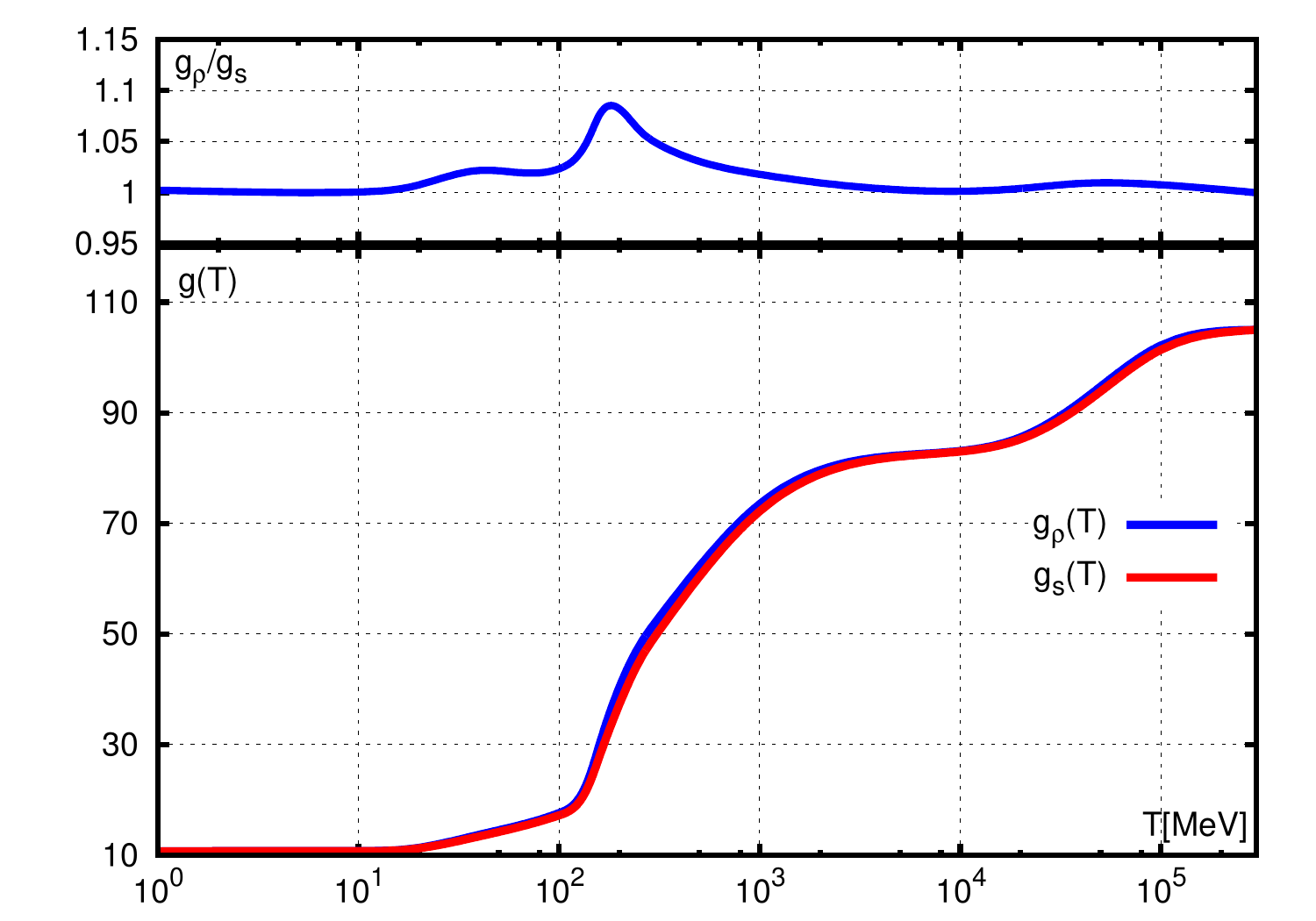}}
    \caption{\label{fig:EOS} The effective number of degrees of freedom $g_*$ as a function of temperature, from Ref.~\cite{Borsanyi:2016ksw}.
    The number depends slightly on whether one determines it from the energy density (blue) or the entropy density (red).
    See the reference for more discussion.}
\end{figure}

The $a q/f_a$ coupling is responsible for the way that the axion solves the strong time-reversal breaking problem.
It sets the mass of the axion in terms of the topological charge density of QCD
$\chitop \simeq (76\:\text{MeV})^4$
\cite{GrillidiCortona:2015jxo,Gorghetto:2018ocs},
namely $m_a^2 = \chitop/f_a^2$.
It also makes it possible for the axion to be generated in the early Universe.
If the axion thermalized in the early Universe at some temperature $T$, when there were $g_*(T)$ Standard-Model thermal degrees of freedom, then it represented one additional degree of freedom at that time.
If it fell out of thermal contact at a decoupling temperature $T_{\text{dc}}$, then we would observe $1+1/g_*(T_{\text{dc}})$ times the expected amount of radiation in the modern Universe.
(For historical reasons cosmologists describe this as adding $43/(7g_*)$ effective neutrino species.)
The number $g_*$ drops significantly near the QCD crossover region, $T \in [150\,\text{MeV},300\,\text{MeV}]$,
see Figure 1 of Ref.~\cite{Borsanyi:2016ksw}, which we reproduce in Fig.~\ref{fig:EOS}.

The most up-to-date constraints on the number of extra radiation species \cite{Goldstein:2026iuu} constrain the number of extra neutrino species to be less than 0.107, implying that $g_* > 57$ when axions decouple.
Future experiments aim to tighten the constraint by a further factor of 3 \cite{CMB-S4:2026mge}, which would constrain or detect the axion decoupling temperature across the whole range shown in Figure \ref{fig:EOS}.
To turn such results into constraints or measurements of the axion coupling-strength and mass, we need to determine the thermal axion production rate as a function of temperature.
Since the range $200\:\text{MeV}<T<1\:\text{GeV}$ is where $g_*$ changes particularly quickly and covers the range where current observational constraints come into effect, it is particularly important to understand the axion production rate in this temperature range.

It is well understood how to relate the axion production rate to correlation functions of the topological charge.  Because axions are extremely light and satisfy the lightlike dispersion relation $k^0 = |k|$, the production rate of axions of momentum $k$ is given by
\begin{align}
\label{BasicRate}
    \odv{\Gamma}{^3 \bm k}
    &= - \frac{1}{(2\pi)^3 2 k^0 f_a^2} \, \frac{\sigma(k,k)}{e^{\beta k^0} - 1} ,
\end{align}
which is correct to first order in the axion-gluon coupling $f_a^{-1}$ and to all orders in the QCD coupling.
Here $\beta = 1/T$ as usual and $\sigma$ is the spectral function of the topological charge density,
\begin{align}
\label{defsigma}
    \sigma(k^0, k)
    &\equiv \int \diff^4{x} \: e^{i(k^0 x^0 - \bm k \cdot \bm x)} \langle \left[ q(x), q(0) \right] \rangleBeta ,
\end{align}
with $\langle \ldots \rangleBeta$ the Minkowski-space thermal expectation value.

The problem is that, while the spectral function $\sigma(k,k)$ can be computed at very high temperatures within perturbation theory
\cite{Bouzoud:2024bom,Bouzoud:2026rur}
and below the transition region in terms of pion scattering
\cite{DEramo:2021psx,Notari:2022ffe},
the transition region itself is precisely where QCD is too strongly coupled to rely on either technique.

Instead, we would like to compute, or at least strongly constrain, the axion production rate using a fully nonperturbative technique which is close to first-principles QCD:  lattice QCD \cite{gattringer-lattice-qcd}.
Lattice QCD only addresses Euclidean correlation functions and \Eqref{BasicRate} involves the Minkowski-domain spectral function.
Because of this, previous approaches to this problem
\cite{Altenkort:2022yhb,Bonanno:2023thi,Bonanno:2026ocw}
have relied on the analytical continuation of Euclidean data to the Minkowski domain, an approach with potentially large systematic errors depending on the reliability of assumptions made in carrying out the continuation.
Here we will present a new method, which uses ideas from recent work by Harvey Meyer and collaborators
\cite{meyer-photon-2018,meyer-photon-2024},
who have shown how \textsl{weighted integrals over} the photon production rate can be computed via Euclidean simulations \textsl{without the need for analytic continuation}.
Following their lead, Section \ref{sec:axioncorr} will present a similar approach for determining weighted integrals over the axion production rate from \textsl{finite-Euclidean frequency, imaginary-spatial-frequency} Euclidean correlation functions.
As we will show, we will be interested in Euclidean-space correlation functions of the form
\begin{align}
\label{Hdefined}
    \HE(\omega_n) & \equiv \int_0^\beta \diff\tau \int \diff^3{x} \:
    \frac{\cos(\omega_n \tau) \sinh(\omega_n |\bm x|)}{\omega_n |\bm x|} \langle q(\bm x,\tau) q(0) \rangleE,
    \\
    \label{omegavalues}
    \omega_n & = 2\pi T \, n, \quad
    n = 0,1,2,\ldots
\end{align}
with $\langle \ldots \rangleE$ the Euclidean-time thermal expectation value.
The challenge in this expression is that the weight function $\propto \sinh(\omega_n |\bm x|)$ grows exponentially with distance, and while 
$\int_0^\beta \diff\tau \: \cos(\omega_n \tau) \langle q(\bm x,\tau) q(0) \rangle$ is guaranteed to decay faster, the statistical fluctuations in its evaluation generically will not.
We will show how to overcome this problem in Section \ref{sec:numerics}, using a combination of gradient-flow techniques
\cite{gradient1-Lüscher2010,Lüscher2010-gradientflow2}
and tail-fitting of the large-separation data.
We will test these techniques for the simplifying case of the pure-glue theory of QCD at a single lattice spacing with 16 sites across the time direction and at a single temperature $T = 1.5 T_c$ as a proof of principle.
Our results are presented in Section \ref{sec:results}, where we show that $H(\omega_n)$ for $n=0$ and $n=1$ can be evaluated with good signal to noise and $n=2$ can be realized with marginal signal to noise with a relatively modest total numerical effort.
We also compare our method with perturbative results from the literature.
Finally, Section \ref{sec:discussion} discusses the implications of our results and what is needed to extend them to full QCD.

\section{Axion production and Euclidean correlation functions}
\label{sec:axioncorr}

We will investigate the production rate of axions via the $a\, q/f_a$ interaction of \Eqref{axionLagrangian}.
Because other, model dependent couplings to axions are possible (For a review see Ref.~\cite{Kim:2008hd}), the axion production rate derived from this interaction is a \textsl{lower bound} on the total axion production rate.
This is sufficient for establishing a lower bound on the coupling strength $1/f_a$, since additional production channels will lead to later axion decoupling and therefore more axions.

\subsection{Axion production from the spectral function}
\label{subsec:axiontospectral}

For completeness we sketch the well-known derivation of the axion production rate in terms of the $q$-$q$ spectral function.
This is analogous to, and in fact simpler than, the production rate for photons from a quark-gluon plasma, which appears in standard field-theory textbooks
\cite{Le-Bellac,Kapusta:2006pm}.
The $S$-matrix element for an initial QCD state $i$ to transition to a final QCD state $f$ together with an axion with momentum $k$ is
\begin{equation}
\begin{aligned}
    S_{fi}(k)
    &= \bra{f,k} T \exp\left[ -i \int \diff^4{x} \, f_a^{-1} a(x) q(x) \right] \ket{i} \\
    &\approx -i f_a^{-1} \int \diff^4{x} \, \bra{f,k} a(x) q(x) \ket{i} = -i f_a^{-1} \int \diff^4{x} \, e^{ikx} \bra{f} q(x) \ket{i} ,
\end{aligned}
\end{equation}
which, squaring, integrating over the final state and tracing over all initial states with a thermal density matrix, becomes
\begin{equation}
\begin{aligned}
    \frac{1}{Z(\beta)} \sum_{if} e^{-\beta E_i} \left| S_{fi}(k) \right|^2
    &=
    \frac{1}{f_a^2 Z(\beta)} \sum_{if} e^{-\beta E_i} \int \diff^4{x} \diff^4{y} \, e^{ik(x-y)} \bra{i} {-}q(y) \ket{f} \bra{f} q(x) \ket{i} \\
    &=
    -f_a^{-2} \int \diff^4{x} \diff^4{y} \, e^{ik(x-y)} \left< q(y) q(x) \right>_\beta 
    =
    -f_a^{-2} V_4 \, G^<(k)
    .
\end{aligned}
\end{equation}
Here $G^<(k)$ is the Wightman function and $V_4$ is the space-time volume.  The minus sign arises because $q$ is antihermitian.
The total axion production rate per unit volume and time is then determined by integrating over the light-cone $|\bm{k}|=k^0$:
\begin{equation}
    \Gamma_\text{tot}
    =
    -f_a^{-2} \int \frac{\diff^3{k}}{(2\pi)^3 2 k^0} \, G^<(k,k) 
\end{equation}
or equivalently
\begin{equation}
\label{eq:diffGamma}
    \frac{\diff \Gamma}{\diff^3{\bm{k}}} =
    - \frac{1}{(2\pi)^3 2k^0 f_a^2} G^<(k,k)
    = - \frac{1}{(2\pi)^3 2k^0 f_a^2}
    \frac{\sigma(k,k)}{e^{\beta k^0}-1} ,
\end{equation}
where we have used the KMS relation \cite{Kubo:1957mj,Martin:1959jp} between the Wightman function and the spectral function introduced in \Eqref{defsigma}.

\subsection{From spectral function to Euclidean function}

\begin{figure}[tb]
\centerline{\includegraphics[width=0.65\textwidth]{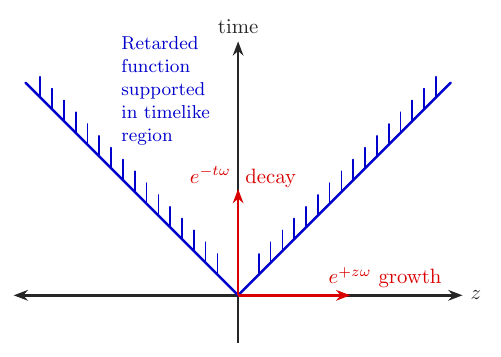}}
    \label{fig:retarded}
    \caption{Retarded function at imaginary frequency and momentum $\Gret(i\omega_n,i\omega_n \hat{k})=\HE(\omega_n)$ is an integral over space and time with an exponentially decaying factor in the time direction and an exponentially growing factor in one space direction.
    Because the retarded function is supported on future timelike or null points $t\geq |\bm{x}|$, the product of the exponentials is always $\leq 1$.
    For the same reason, $\Gret(i\omega_n,\bm{x})$ involves an integral over time $t\geq |\bm{x}|$ with a factor $e^{-\omega_n t}$ and therefore decays in space at least as fast than $e^{-\omega_n |\bm{x}|}$.}
\end{figure}

Next, we relate the light-cone spectral function to an Euclidean time correlator.
For this we follow a technique first introduced in Ref. \cite{meyer-photon-2018} in the context of the photon production rate.
Consider the retarded correlator at light-like momentum with the energy $\omega$ considered as a complex variable,
\begin{equation}
\label{Hrdef}
\begin{aligned}
    \Hret(\omega)
    \equiv
    \Gret(\omega, \omega \hat{k}) 
    =
    i \int \diff^4{x} \, e^{i\omega (x^0 - \hat{k} \cdot \bm{x})} \, \theta(x^0) \left\langle [q(x), q(0)] \right\rangle_{_{\!\beta}}
    ,
\end{aligned}
\end{equation}
were $\hat{k}$ denotes some arbitrary unit spatial direction.
Its imaginary part is given by the spectral function, $\Im \Hret(\omega) = 2 \sigma(\omega, \omega)$, and may be related to its real part by use of the Kramers-Kronig relation,
\begin{equation}
    \Re \Hret(\omega)
    =
    P \int_{0}^\infty \frac{\diff{\omega'}}{\pi} \, \omega' \, \frac{\sigma(\omega', \omega')}{\omega'^2 - \omega^2}
    ,
\label{eq:KramersK}
\end{equation}
under two assumptions.
First, $\Hret(\omega)$ must be analytic in the upper half plane.
To show this, it suffices to show that the exponential has a decaying part.
Due to causality, the retarded function is non-zero only if $x$ is future timelike or null, $x^2 = -(x^0)^2 + |\bm{x}|^2 \leq 0$, see Figure~\ref{fig:retarded}.
Since $\hat{k}$ is a unit vector we have $(x^0)^2 \geq |\bm{x}|^2 \geq (\hat{k} \cdot \bm{x})^2$, which together with $x^0>0$ gives us $x^0 - \hat{k} \cdot \bm{x} \geq 0$.
Thus, for $\Im \omega > 0$, the exponential is less than 1 and $\Hret(\omega)$ is analytic.
The second assumption is that $\omega \Hret(\omega)$ must go to zero as $|\omega|$ goes to infinity.
If this condition is not satisfied, then one can instead consider appropriate differences of form $\Hret(\omega_1)-\Hret(\omega_2)$, for which \Eqref{Hrdef} contains a difference of two complex exponentials and \Eqref{eq:KramersK} contains a difference of $1/(\omega'{}^2-\omega^2)$ factors which leads to a convergent integral.
Later we will see that such differencing is both necessary and sufficient for our case.

To make the connection to Euclidean time, we use that the retarded correlator is related to the Euclidean correlator through analytic continuation,
\begin{equation}
    \GEucl(\omega_n, \bm{k})
    = \Gret(i\omega_n, \bm{k})
    \quad \forall n \in \{ 0,1,2,\ldots \}
    ,
\end{equation}
with the Euclidean correlator defined as
\begin{equation}
    \GEucl(\omega_n, \bm{k})
    \equiv \int_0^\beta \diff{\tau} e^{i\omega_n \tau} \int \diff^3{\bm{x}} e^{i \bm{k} \cdot \bm{x}} \langle q(x) q(0) \rangleE
    .
\label{eq:GEdef}
\end{equation}
By taking the spatial momentum to be imaginary we can define a Euclidean light-like correlator,
\begin{equation}
    \HE(\omega_n)
    \equiv \GEucl(\omega_n, i\omega_n \hat{k})
    = \Gret(i\omega_n, i\omega_n \hat{k})
    = \Hret(i\omega_n)
    .
\label{eq:eucl-ll}
\end{equation}
Inserting \Eqref{eq:eucl-ll} into \Eqref{eq:KramersK}, and noting that $\GEucl$ is a real function, we obtain
\begin{equation}
    \HE(\omega_n)
    = \int_0^\infty \frac{\diff{\omega'}}{\pi} \omega' \frac{\sigma(\omega', \omega')}{\omega'^2 + \omega_n^2}
    = -2\pi f_a^2 \int_0^\infty \frac{\diff{\omega'}}{n_b(\omega') (\omega'{}^2+\omega_n^2)} \frac{\diff\Gamma}{\diff\omega'}.
\label{eq:nosubyet}
\end{equation}
The left side is what we will compute on the lattice.
Due to the imaginary spatial momentum, the second exponential in \Eqref{eq:GEdef} becomes purely real, thus enhancing the integrand exponentially in the direction of $\hat{k}$, and suppressing it in the opposite direction.
This results in a rapid decay of the signal-to-noise ratio (SNR) at large distances, whose solution in terms of fits we will discuss in Section \ref{sec:numerics}.
We can slightly improve the situation by using rotational symmetry by averaging over $\hat{k}$, resulting in 
\begin{equation}
\label{HErotated}
    \HE(\omega_n)
    = \int_0^\beta \diff\tau \int \diff^3{x} \:
    \frac{\cos(\omega_n \tau) \sinh(\omega_n |\bm x|)}{\omega_n |\bm x|} \langle q(\bm x,\tau) q(0) \rangleE
    .
\end{equation}
For the case $n=0$, $\omega_n=0$ and the $\omega_n$-dependent ratio above is just 1.

\subsection{Subtractions}

\begin{figure}[h]
    \centering
    \includegraphics[width=0.75\linewidth]{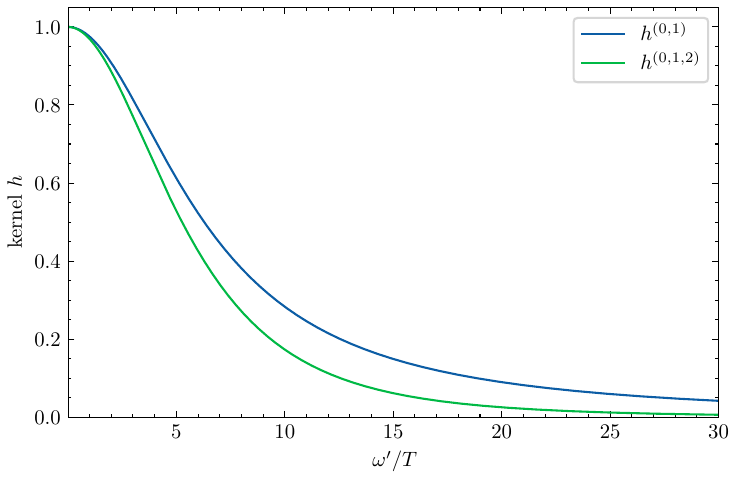}
    \caption{The kernel functions $h^{(0,1)}$ and $h^{(0,1,2)}$.
    }
    \label{fig:h-plot}
\end{figure}

The derivation of \Eqref{eq:nosubyet} involved a contour rotation which only works if the integral on the RHS of that equation is convergent.
We will show below that it is not convergent, which means that the expression misses a contribution from the large-complex-frequency contour.
However, if we consider instead linear combinations of $\HE(\omega_n)$ functions such that the large-$\omega$ behavior shows better convergence, then the expression becomes valid.
Therefore we consider differences between two or more Matsubara modes,
\begin{align}
    \HE^{(n_0, n_1, \ldots)}
    & \equiv \sum_i c_i \HE(\omega_{n_i})
    =
    \int_0^\infty \frac{\diff{\omega'}}{\pi} \frac{\sigma(\omega', \omega')}{\omega'} h^{(n_0, n_1, \ldots)}(\omega')
    , \\
    h^{(n_0, n_1, \ldots)}(\omega')
    & =
    \sum_i c_i \frac{ \omega'^2}{ \omega'^2 + \omega_{n_i}^2}
    ,
\label{eq:H-diffs}
\end{align}
where the coefficients $c_i$ are to be chosen in order to cancel the slowest-decaying powers of $\omega'$.
Combining the two and the three lowest Matsubara modes, we propose the combinations
\begin{align}
    \HE^{(0,1)}
    &=
    \HE(\omega_0) - \HE(\omega_1) =
    \int_0^\infty \frac{\diff{\omega}'}{\pi}
    \frac{\sigma(\omega', \omega')}{\omega'}
    \frac{(2\pi T)^2}{\omega'{}^2 {+} (2\pi T)^2}
    ,
    \label{eq:sub-01}\\
    \HE^{(0,1,2)}
    &=
    \frac{3\HE(\omega_0)-4 \HE(\omega_1)+\HE(\omega_2)}{3}
    =  \int_0^\infty \frac{\diff{\omega}'}{\pi}
    \frac{\sigma(\omega',\omega')}{\omega'}
    \frac{(2\pi T)^2(4\pi T)^2}{(\omega'{}^2 {+} (2\pi T)^2)(\omega'{}^2 {+} (4\pi T)^2)}
    . \label{eq:sub-012}
\end{align}
We see that these functions converge by 2 and 4 more powers of $\omega'{}^{-1}$ respectively, which will be enough to render the integrals finite.
The functions $h^{(0,1)}$ and $h^{(0,1,2)}$ are plotted in Fig. \ref{fig:h-plot}.

\subsection{High-frequency and short-distance behavior}
\label{subsec:shortdist}

Next we check whether the large-frequency behavior of the integrals \Eqref{eq:nosubyet}, \Eqref{eq:sub-01} and \Eqref{eq:sub-012} are sufficiently convergent.
First consider the frequency domain.
On dimensional grounds $\sigma(\omega, k)$ has energy-dimension 4.
At vanishing spatial momentum we would then find $\sigma(\omega, 0) \propto \omega^4$, which would make \Eqref{eq:nosubyet} severely large-$\omega'$ divergent.
But on the light-cone, the \textsl{vacuum} value of $\sigma(\omega', \omega')$ vanishes -- since it is equivalent under a boost to the value with $\omega'$ arbitrarily small.
Perturbatively, the thermal contribution at large frequency scales as $\sigma(\omega',\omega') \propto \omega' T^3$ up to logs \cite{Bouzoud:2026rur}, making the individual $\HE(\omega_n)$ divergent but differences such as \Eqref{eq:sub-01} safely large-$\omega'$ convergent.

One can also consider the spacetime domain and ask about the short-distance behavior of \Eqref{HErotated} and its subtracted versions.
First consider $H^0$.
The $\langle q(r,\tau) q(0,0)\rangleE$ correlation function diverges at short distances as
\begin{align}
\label{eq2.1}
- \langle q(r,\tau) q(0,0) \rangleE & \to_{r,\tau \to 0} \frac{3(N_c^2-1) g^4}{64\pi^8} \frac{1}{(r^2+\tau^2)^4},
\end{align}
with $N_c=3$ the number of colors.
This renders \Eqref{HErotated} for $\omega_n=0$,
\begin{align}
\HE(0) = \int \langle q(r,\tau) q(0) \rangleE \: \diff^3 r \diff\tau,
\end{align}
severely short-distance divergent.
The integral over this divergent behavior is canceled by a contact term at precisely zero separation.
Performing the integral including this contact term results in $\HE(0) = \chitop$ the topological susceptibility.
This finite result arises from the cancellation of two formally divergent quantities.
Correspondingly, the interpretation of \Eqref{Hdefined} requires a UV (large-momentum) subtraction.

However, if we instead consider the difference
\begin{align}
\label{eq:H01est}
\HE^{(0,1)} & 
=
\int \diff^3{r} \diff{\tau} \: \langle q(r,\tau) q(0,0) \rangleE
\left[ 1 - \cos(\omega_1 \tau) \frac{\sinh(\omega_1 r)}{\omega_1 r} \right], & \omega_1 & = 2\pi T\,,
\end{align}
the term in square brackets goes to 0 as $r,\tau \to 0$.
This eliminates the contact term and makes the short-distance divergence less severe.
In fact, expanding the quantity in brackets about small separation, we find
\begin{align}
\label{eq:angleavg}
\left[ 1 - \cos(\omega_1 \tau) \frac{\sinh(\omega_1 r)}{\omega_1 r} \right]
\simeq \omega_1^2 \left( \frac{\tau^2}{2} - \frac{r^2}{6} \right),
\end{align}
which averages to zero over any small 3-sphere $S^3$ at fixed $r^2+\tau^2$-value (whether or not we expand it in small separation).
The leading divergence in Eq.~(\ref{eq2.1}) is $\mathrm{SO}(4)$ invariant, and provided that one carries out this short-distance direction-averaging first, it will therefore cancel.
Any $\text{SO}(4)$ non-invariant contribution to Eq.~(\ref{eq2.1}) must arise as a thermal effect.
At short distances we can expand Eq.~(\ref{eq2.1}) through an operator product expansion, and since the lowest-dimension operator available%
\footnote{Caron-Huot used similar arguments in \cite{Caron-Huot:2009ypo} when analyzing stress-tensor correlation functions.}
has dimension $\Delta =4$, the first rotation-non-invariant thermal corrections scale at worst as $T^4/r^4$.
Inserted into the integral of Eq.~(\ref{eq:H01est}), this will give rise to a short-distance finite result.
This explains why the integral in \Eqref{eq:sub-01} is large-$k$ convergent.
The same holds for any expression in the form of \Eqref{eq:H-diffs} provided that $\sum_i c_i = 0$.

We should also check that the large-distance part of the integral converges.
The large-distance behavior of $\HE(0)$ is finite because $\langle q(x) q(0) \rangleE$ decays exponentially at large separation with a mass controlled by the lightest flavorless $0^{-+}$ state (further explained in Section \ref{sec:fitfcn}).
For $\HE(\omega_n)$ we can first carry out the $\int \diff{\tau} \, \cos(\omega_n \tau)$ to obtain the $n$'th Matsubara mode, which is guaranteed to decay at large separation faster than $\exp(-\omega_n r)$ \cite{Caron-Huot:2022akb}, as we saw in Figure \ref{fig:retarded}.
This is sufficient to prevent large-distance issues in any $\HE(\omega_n)$.

Therefore, if we can compute more than one of the $\HE(\omega_n)$ with adequate SNR, we can combine them to generate finite weighted integrals of the axion production rate.

\section{Lattice calculational approach}
\label{sec:approach}

The purpose of this paper is to carry out a first exploratory study to examine the feasibility of computing $\HE^{(0,1)}$ and $\HE^{(0,1,2)}$.
Therefore we will work within pure $SU(3)$ Yang-Mills theory without fermions (equivalent to the quenched approximation).
We will also only investigate a single temperature, $T = 1.5 T_c$, and a single lattice spacing, with $N_\tau = 16$ sites in the time direction, so our lattice spacing is $a = 1/(16T)$.
Based on community experience and the very rapid decay of $0^-$ correlation functions, it is sufficient to consider an aspect ratio of 4, that is, $N_s = 64$ lattice spacings in the three spatial directions.
We use the Wilson gauge action, for which the lattice coupling at our desired temperature is $\beta=6.868$
\cite{scale-set-r0Tc,scale-set-r0Tc-upd}.
We generate $N=4000$ configurations, each separated by 100 combined sweeps, with one combined sweep consisting of four over-relaxation sweeps followed by one heat bath sweep.
We have checked that this is sufficient to remove autocorrelations.
To ensure thermalization we discard the first 5500 combined sweeps.
All computations described in this section are performed within the MILC code base \cite{milc} on the Lichtenberg high-preformance computing center%
\footnote{\texttt{https://www.hrz.tu-darmstadt.de/hlr/hochleistungsrechnen/index.en.jsp}}%
, and they represent about 60,000 CPU core-hours (which by modern standards is a very modest numerical effort).

\begin{figure}[tb]
    \centering
    \includegraphics[width=0.9\linewidth]{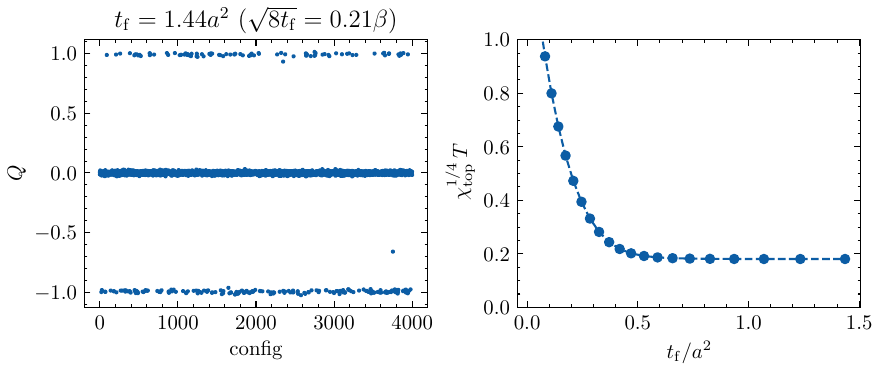}
    \caption{Left: the topological charge takes integer values when sufficient gradient flow is applied.
    Right: topological susceptibility as a function of the flow time, showing that zero or small flow times lead to spurious fluctuations in $\chitop \equiv \langle Q^2 \rangleE/V_4$.}
    \label{fig:suscept}
\end{figure}

We use the gluonic definition (\Eqref{qdef}) for the topological charge density on the lattice, with an $a^4$-improved field strength tensor.
The field strength is built out of normal clover terms made out of square plaquettes, mixed with clover terms made of 1x2 rectangular plaquettes to remove $a^2$ errors \cite{impr-fs2}.
This definition of $q$ suffers severe multiplicative renormalization and contamination with other operators unless we evaluate it on gradient-flowed gauge fields
\cite{gradient1-Lüscher2010,Lüscher2010-gradientflow2},
which filters out the short-distance fluctuations which are responsible for these problems.
Specifically, we have chosen to use improved (Zeuthen) gradient flow \cite{zeuthen-flow}.
Gradient flow introduces a new parameter $t_{\text f}$, called the flow time%
\footnote{note that $\tflow$ has units of length-squared.},
along which configurations are flown in the direction of decreasing action.
We illustrate the importance of gradient flow in Figure \ref{fig:suscept}, which shows how the total topology $Q \equiv \int q \diff^4{x}$ correctly takes an integer value when we apply sufficient flow, and how the topological susceptibility $\chitop \equiv \langle Q^2 \rangleE / V_4$ exhibits spurious contributions at zero and small flow times but stabilizes by $\tflow \geq 0.4 a^2$.
The disadvantage of applying gradient flow is that $q(x)$ must be interpreted as an average over a region around $x$ with radius $r \sim \sqrt{8\tflow}$, rather than the value precisely at the point $x$.
This can change the result of integrals like \Eqref{HErotated}, and it means that we will have to extrapolate our results to zero flow time, which we will do as a final step in our analysis.

To compute the euclidean correlator in \Eqref{HErotated} we need the charge correlators for all separations,
\begin{equation}
    \Gtf(x)
    \equiv \frac{1}{\Vlatt}  \sum_{y \in \Lambda} q(y+x) q(y) \Big|_{t_{\text f}} \quad \forall x \in \Lambda
    ,
\end{equation}
where we average over the whole lattice $\Lambda$ with $\Vlatt$ denoting the number of lattice sites.
Naively, this requires $\mathcal{O}(\Vlatt^2)$ operations but, as shown in Ref. \cite{fourier-trick-1}, the use of the Fast Fourier Transform allows for an $\mathcal{O}(\Vlatt \log \Vlatt)$ computation.
Next, we average $\Gtf(x)=\Gtf(\tau, \bm{x})$ over all $\bm{x}$-values with the same magnitude $r$, and perform a discrete Fourier transform in the temporal direction,
\begin{equation}
    \Gtf(\omega_n, r)
    = \sum_{\tau=0}^{N_\tau - 1} e^{i \omega_n \tau} \left( \frac{1}{d_r} \sum_{|\bm{x}|=r} \Gtf(\tau, \bm{x}) \right)
    .
\label{eq:corrlat}
\end{equation}
Here, $d_r$ is the number of lattice sites in a 3D spatial slice with spatial distance $r$ from the origin
(OEIS%
\footnote{\texttt{https://oeis.org/A005875}}
series \texttt{A005875}).
Note that this expression breaks down at large distances $r \ge L/2a$ with $L$ the spatial length of the lattice and $a$ the lattice spacing.
Our lattice is large enough that $\Gtf(\omega_n,L/2a)$ is consistent with zero, and if it were not, then we would need a larger lattice.

The function $\Gtf(\omega_n,r)$ is what we need to compute \Eqref{HErotated}, but there will be some subtleties which we will address next.

\section{Numerical details}
\label{sec:numerics}

We turn now to the computation of the integral in \Eqref{HErotated} which on the lattice is given by
\begin{equation}
    \label{eq:latticegoal}
    \HEtf(\omega_n)
    = \sum_r d_r \frac{\sinh(\omega_n r)}{\omega_n r} \Gtf(\omega_n, r)
    ,
\end{equation}
with $\Gtf(\omega_n, r)$ the charge correlators obtained from the lattice as described in the last section, which is a function of flow time $\tflow$, Matsubara frequency $\omega_n$ and spatial separation $r$.
In practice, a straightforward evaluation is impossible since the exponential factor causes the SNR of the integrand to grow exponentially with distance $r$.
The solution is to replace the large-$r$ part of  $\Gtf(\omega_n, r)$ with a fit.
First, we derive the fitting function and discuss its parameters.
We then perform the fit and discuss the problem of strong correlations in the data.
Having the fits we compute the integrals, and finally perform the zero flow-time extrapolation.

\subsection{The fitting function}
\label{sec:fitfcn}

We will fit $\Gtf(\omega_n,r) \propto \sum_i c_i e^{-m_i r}/r$ with coefficients $c_i$ and masses $m_i$ treated as fitting parameters.
Let us explain briefly the motivation for this fitting form.
Following the logic of Ref.~\cite{Arnold:1995bh}, consider the $q-q$ correlation function integrated over a transverse $x,y$ plane as a function of $z$:
\begin{equation}
\begin{aligned}
    &\quad \beta V_\perp G(p_\tau = \omega_n, p_\perp = 0, z) \\
    &= \left< \left( \int \diff{\tau} e^{-i\omega_n \tau} \int \diff^2{x_\perp} q(\tau, x_\perp, z) \right) \left( \int \diff{\tau'} e^{i\omega_n \tau'} \int \diff^2{x_\perp'} q(\tau', x_\perp', 0) \right) \right>
    .
\end{aligned}
\label{eq:xy-0p}
\end{equation}
Here, $x_\perp$ and $p_\perp$ denote position and momentum in the $xy$-plane, respectively, and $V_\perp$ its area.
We now reinterpret the $\tau$-direction as a compact spatial direction, and $z$ as the temporal direction.
Since the $z$-extent is infinite, this reinterpretation results in a zero-temperature theory in $\mathbb{S} \times \mathbb{R}^2$ space.
In this theory, the Hamiltonian $H$ generates translations in $z$, and we can write
\begin{align}
\label{eq:xy-0p-2}
    \beta V_\perp G(p_\tau = \omega_n, p_\perp = 0, z)
    & = - \bra{0} q_n^\dagger e^{-H z} q_n \ket{0}, 
    \\ \nonumber
    q_n & \equiv \int_0^\beta \diff{\tau} e^{i \omega_n \tau} \int \diff^2 x_\perp q(\tau, x_\perp, 0)
    .
\end{align}
The minus sign arises from the non-hermiticity of the charge density $q(x)$.
Acting on the vacuum, the operator $q_n$ excites certain states in the $n$-th Matsubara sector, $q_n \ket{0} = \sum_\nu c_n^{\nu} \ket{n,\nu}$, with energies $H \ket{n,\nu} = m_n^\nu \ket{n,\nu}$.
Substituting into \Eqref{eq:xy-0p-2}, we obtain the spectral representation
\begin{equation}
    \beta V_\perp G(p_\tau = \omega_n, p_\perp = 0, z)
    = \int \diff{m} \, \sigma_n(m) e^{-mz}
    , \quad
    \sigma_n(m)
    \equiv - \sum_\nu |c_n^\nu|^2 \delta(m - m_n^\nu)
    .
\label{eq:def-sigma-m}
\end{equation}
Next, we assume that the low-mass part of the spectral function $\sigma_n(m)$ is supported on discrete states without continuum contributions, which we expect to be true in an interacting theory like QCD.%
\footnote{In vacuum this would be true up to the $0^{-+} 0^{++}$ two-particle threshold.
An analysis of the glueball spectrum finds at least two states below this threshold \cite{glueballs-2020}.}
Next, we Fourier transform in $z$ and then use rotational invariance in $xyz$-space to obtain
\begin{equation}
    \quad \beta V_\perp G(\omega_n,\bm{p})
    = \int \diff{m} \, 2 m \, \sigma_n(m) \frac{1}{m^2 + |\bm{p}|^2}
    ,
\end{equation}
followed by a Fourier transform in $\bm{p}$, giving
\begin{equation}
    \beta V_\perp G(\omega_n, \bm{x})
    = \int \diff{m} \, \frac{m \, \sigma_n(m)}{2 \pi} \frac{e^{-m |\bm{x}|}}{|\bm{x}|}
    .
\end{equation}
Inserting back the definition of $\sigma_n(m)$ from \Eqref{eq:def-sigma-m}, we obtain our fitting function,
\begin{equation}
    \Gtf^\text{fit}(\omega_n, r)
    = \sum_\nu A_n^\nu(\tflow) \frac{e^{-m_n^\nu r}}{r}
    , \qquad
    A_n^\nu(\tflow)
    \equiv - \frac{1}{2 \pi \beta V_\perp} m_n^\nu \, |c_n^\nu(\tflow)|^2
    .
\label{eq:fitf-gen}
\end{equation}
The fitting parameters are the masses $m_n^\nu$ and the amplitudes $A_n^\nu(\tflow)$.
Since we only want to fit the tail which is controlled by the smallest masses appearing in the sum, we will consider either one or two masses.

In the last step above we reintroduced the dependence on gradient flow depth $\tflow$, but this point really deserves a little more discussion.
The operator $q(\tflow,\tau,x_\perp,0)$ actually extends a short distance $r_{\text f} \sim \sqrt{\tflow}$ in the $z$-direction, which complicates the Hamiltonian interpretation of \Eqref{eq:xy-0p-2}.
But for $z^2 \gg \tflow$, the two operators in \Eqref{eq:xy-0p-2} will be in the correct $z$-ordering up to exponentially suppressed corrections, and the derivation goes through up to similarly exponentially-suppressed corrections.
However, the $q_n(\tflow)$ operator is really a different operator at different $\tflow$ values, so the amplitudes with which it creates different states will be $\tflow$ dependent, and so will the coefficients $A_n^\nu(\tflow)$.
However, the masses correspond to the physical masses in the $\mathbb{S} \times \mathbb{R}^2$ theory. 
They should \textsl{not} depend on the choice of flow time, provided that we only use data with $r^2 \gg \tflow$.

To get a first estimate for the masses we can compare to studies of glueball masses in the vacuum, for example Ref.~\cite{glueballs-2020}.
Since in our case the masses correspond to a theory with one periodic dimension, we do not expect them to match exactly.
Looking at \Eqref{eq:xy-0p-2}, the states excited by $q_0$ are zero-spin zero-momentum states, odd under reflection in space and time, and even under charge conjugation.
From \cite{glueballs-2020} the first two such $J^{PC}=0^{-+}$ glueball states have continuum-extrapolated masses of $m \in \{6.24T, 8.8T\}$, where $T=1.5 T_c$.

For masses in higher Matsubara sectors (those excited by $q_n$ with $n>0$), we have the constraint $m_n > 2\pi n T$ \cite{Caron-Huot:2022akb} as mentioned previously.
As a first guess we may use the zero-temperature continuum energy-momentum relation $m_n = \sqrt{\omega_n^2 + m_0^2}$, although we do not expect this guess to be good due to the presence of one periodic dimension from the finite temperature.

\subsection{The fitting procedure}

\begin{figure}[tb]
    \centering
    \includegraphics[width=0.98\linewidth]{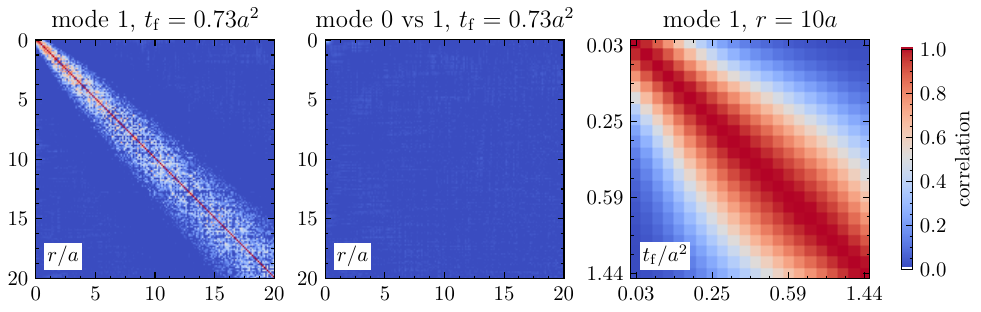}
    \caption{The degree of correlation between different radii at a fixed gradient-flow depth (left), between different Matsubara modes (middle), and between different gradient flow depths at a fixed radius and $\tflow$ (right).
    Both radius and gradient flow show correlations between nearby values, but the different Matsubara modes are not correlated.}
    \label{fig:correlations}
\end{figure}

We want to fit the large-$r$ tail of the data $\Gtf(\omega_n,r)$ to the fitting function \Eqref{eq:fitf-gen}.
Since the masses do not depend on the flow time $\tflow$, the data for different flow times should  be fit simultaneously.
Another reason is that the noise in the data is correlated across flow time.
This is because gradient flow suppresses high momentum modes.
Correlators at differing flow times share the low momentum modes and thus also share fluctuation in these modes, leading to positive correlations in $\tflow$.
Similarly, gradient flow acts as smearing in coordinate space, leading to positive correlations in $r$.
Finally, there are no significant correlations between different Matsubara frequencies $\omega_n$.
We illustrate this in Figure \ref{fig:correlations}.
Based on these considerations, we will do a correlated fit of $\Gtf(\omega_n,r)$ simultaneously in $\tflow$ and $r$, but separately for different $\omega_n$.

For the correlated fit we need the covariances in $\tflow$ and $r$ (for each $\omega_n$).
The estimator of the covariance is computed from the lattice data as
\begin{equation}
    \text{Cov}_{\omega_n}(\tflow,r ; \tflow', r')
    \equiv \frac{1}{N(N-1)} \sum_{s=1}^N \left[ \Gtf^{(s)}(\omega_n,r) - \Gtf(\omega_n,r) \right] \left[ G_{\tflow'}^{(s)}(\omega_n,r') - G_{\tflow'}(\omega_n,r') \right]
    .
\label{eq:cov}
\end{equation}
Here, $G^{(s)}$ is the correlator evaluated on the configuration $s$, and the average is over all $N$ configurations.
The problem is that this estimator underestimates small eigenvalues of the covariance matrix unless $N \gg N_D$, where $N_D$ is the number of data points, which is the number of $(\tflow,r)$-pairs in our case.
If an eigenvalue is underestimated, the corresponding mode in the data is given too much weight, potentially leading to poor fits.
For a nice discussion of this problem, see \cite[App. D]{svd-cut-bmeson}.
On our lattice there are $\#r = 1914$ distinct values of $r$, and for a flow-time extrapolation we use at least $\#\tflow = 3$ flow-time values.
This results in $N_D = \#\tflow \cdot \#r \approx 6000 > N$ data points, making the covariance matrix \Eqref{eq:cov} even singular.

Our solution is to bin the data in $r$ which, due to the correlations in $r$, results only in a marginal loss of information.
The idea is to choose the bins such that the error introduced by binning is negligible compared to the uncertainty in the data.
To estimate the binning error, we first do a preliminary fit to the data with constant-size bins, resulting in $G_{\text{fit}_0}(r)$ (suppressing all parameters except $r$).
For some bin $b$, we then define the binning error as
\begin{equation}
    \epsilon_\text{bin}
    = \left| \frac{1}{|b|} \sum_{r\in b} G_{\text{fit}_0}(r) - G_{\text{fit}_0}(r_b) \right|
    , \qquad r_b = \frac{1}{|b|} \sum_{r\in b} r
    .
\end{equation}
The data error is just the uncertainty of the average of all data points contained in $b$,
\begin{equation}
    \epsilon_\text{data}^2
    = \frac{1}{|b|^2} \sum_{r,r' \in b} \text{Cov}[G(r);G(r')]
    .
\end{equation}
Now, bins are chosen as large as possible while obeying
\begin{equation}
    \frac{\epsilon_\text{bin}}{\epsilon_\text{data}} < r_\text{tol}
    ,
\end{equation}
with some binning error tolerance $r_\text{tol} \ll 1$.
For all following fits we will use $r_\text{tol} = 0.01$.
Using this binning procedure, the bins increase in size with increasing distance $r$, reflecting the decrease in SNR for larger $r$.

\begin{table}[tb]
    \centering
    \begin{tabular}{c|ccccccccc}
        \diagbox[width=2.5em,height=2em]{$n$}{$\tflow$} & $0.28a^2$ & $0.37a^2$ & $0.42a^2$ & $0.47a^2$ & $0.59a^2$ & $0.73a^2$ & $0.94a^2$ & $1.23a^2$ & $1.44a^2$ \\
         \hline
        $0$ & 11 & 11.5 & 11.5 & 11.5 & 12 & 12.5 & 12.5 & 13.5 & 14.5 \\
        $1$ & 7.5 & 7.5 & 8 & 8.5 & 8.5 & 9 & 9.5 & 10 & 10.5 \\
        $2$ & 6 & 6 & 6 & 6.3 & 6.3 & - & - & - & - \\ \hline \hline
    \end{tabular}
    \caption{The values of $r_0(n, \tflow) / a$ for the first three Matsubara modes $n$ and a range of flow times $\tflow$, as determined from individual one-mass fits.}
    \label{r0_gs_table}
\end{table}

\begin{figure}[ht]
    \centering
    \includegraphics[width=1\linewidth]{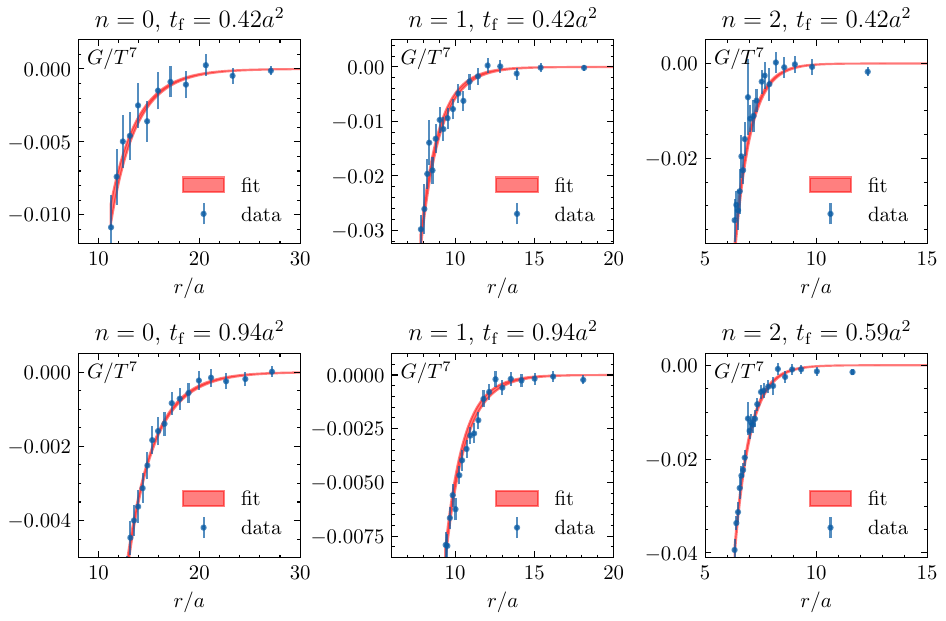}
    \caption{Plots of fit and data for the first three Matsubara sectors and two flow times each.
    The fit includes only the lowest mass for each mode, and is done \emph{simultaneously} over flow times from $t_\mathrm{f}=0.42a^2$ to $t_\mathrm{f}=0.94a^2$ for $n=0,1$, and over flow times from $t_\mathrm{f}=0.42a^2$ to $t_\mathrm{f}=0.59a^2$ for $n=2$.}
    \label{fit-data-plot-gs}
\end{figure}

Next we must determine the range of $r$-values used in the tail fit.
We use $r$-values starting from a value $r_0(\tflow,n)$.
We extend the fit range up to the maximum $r$, but in practice the data beyond some radius are irrelevant in the fit because of their small SNR.
The $r_0$ value is made $\tflow$ dependent because the quality of the data depends strongly on $\tflow$.
We determine $r_0$ by making a mass plot for a single $\tflow$ value and looking for a mass plateau which gives an acceptable quality of the mass fit.
We will consider both a single-mass fit and a two-mass fit; as expected the $r_0$ value providing a good fit is somewhat smaller for the two-mass fit.
The $r_0$ values we use for the one-mass case are shown in Table~\ref{r0_gs_table}.
In the simultaneous fits, we use the flowtimes $t_\mathrm{f}/a^2 \in \{0.42,0.47,0.59 \}$ for the mode $n=2$, and for $n=0,1$ we additionally use $t_\mathrm{f}/a^2 \in \{0.73,0.94 \}$.
To be conservative, we excluded the two largest flowtimes $t_\mathrm{f}/a^2 \in \{1.23, 1.44 \}$, although, in retrospect, they could probably be included as well.
We illustrate the resulting tail fits for the one-mass Ansatz in
Figure~\ref{fit-data-plot-gs}.
As the figure shows, a single-mass fit is an adequate description of the large-$r$ tail at all three $n$-values we consider.
For completeness, we also list the smallest mass in each sector, determined by the 1-mass fit, in Table \ref{masses-table-new}.

\begin{table}[tb]
    \centering
    \begin{tabular}{c|ccc}
         & $m_0 / T$ & $m_1 / T$ & $m_2 / T$ \\
         \hline
        guess & $6.24$ & $8.86 $ & $14.03 $ \\
        one-mass fit & $4.93(24)$ & $11.38(66)$ & $20.24(49)$  \\
    \end{tabular}
    \caption{Lightest mass in each Matsubara mode, compared to our guess based on the vacuum mass spectrum.
    Uncertainties are obtained from 1000 bootstrap samples.}
    \label{masses-table-new}
\end{table}

\subsection{Computing the correlators}

We can finally turn to the evaluation of \Eqref{eq:latticegoal},
which we will slightly rewrite as:
\begin{align}
    \HEtf (\omega_n)
    & =
    \sum_r d_r \hEtf(\omega_n, r)
    , &
    \hEtf(\omega_n, r)
    & \equiv
     \frac{\sinh(\omega_n r)}{\omega_n r} \Gtf(\omega_n, r)
     .
\label{sub-corr-indi}
\end{align}
Following Ref.~\cite{altenkort-blocking}, we use the lattice data at small $r$ where the data is precise and the fitting form is unknown, the fit at large $r$ where the data is noisy, and a linear combination in an intermediate range where the fitting function matches the data.
Specifically, we will evaluate $\HEtf(\omega_n)$ using:
\begin{align}
    \label{eq:whatwedo}
    \HEtf(\omega_n)
    & = \sum_r d_r \big[ (1 - x(r)) \hEtf(\omega_n, r) + x(r) \hEtf^{(\text{fit})}(\omega_n, r) \big]
    , \\ \nonumber
    x(r) & =
    \begin{cases}
    0, & r < r_0, \\
    \frac{r - r_0}{r_1 - r_0}, & r_0 \le r \le r_1, \\
    1, & r_1 < r .
    \end{cases}
\end{align}
The value $r_0$ was already introduced as the minimum radius where the fit is effective.
We choose $r_1$ such that the SNR is poorer than the point $r_0$ by a factor of 5.
In determining the SNR, we use the measured noise but estimate the signal using the fit rather than the data, so that fluctuations in the data do not appear as fluctuations in the SNR.

\begin{figure}
    \centering
   \includegraphics[width=0.99\linewidth]{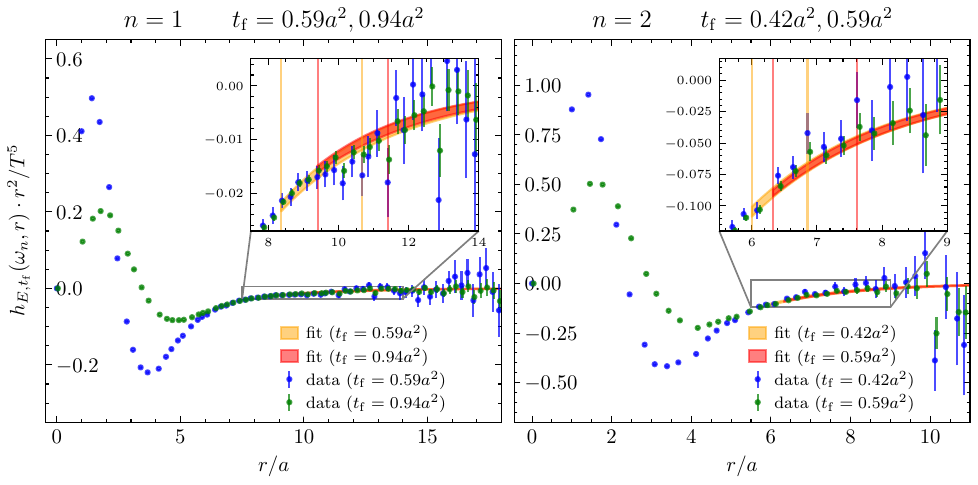}
    \caption{Plot of $r^2 h_{E,\tflow}(r)$ versus $r$ for $n=1,\tflow=0.94a^2$ (left) and $n=2,\tflow=0.57a^2$ (right), illustrating that even with the exponentially growing kernel and a factor $r^2$ to reflect how $d_r$ grows, the peak region in the integrand of \Eqref{sub-corr-indi} is captured by our data.
    The vertical bars indicate the lower and upper regions of our fit range $r_0$ and $r_1$, and the bands show the fit with errors.}
    \label{fig:decays-in-time}
\end{figure}

The central challenge of our approach for $n=1,2$ is that the envelope function $\sinh(\omega_n r)/\omega_n r$ grows exponentially with distance.
We must retain SNR out to a large enough radius that the exponential decay in the data has outrun this growing envelope and the dominant contributions to \Eqref{sub-corr-indi} have already been captured.
This is illustrated in Figure~\ref{fig:decays-in-time}, which shows $r^2 h_{E,\tflow}(r)$ as a function of $r$ for the $n=1$ and $n=2$ cases at the largest flow time we use, together with one slightly smaller flow time.
The figure shows that the most important $r$-values for the integral are successfully captured by the region where we have data with an adequate SNR.
The large wrong-sign contribution at small radius is the result of the contact term discussed in Subsec.~\ref{subsec:shortdist}, smeared by gradient flow.

\subsection{Small-flow-time extrapolation}

In previous investigations using gradient flow to study correlation functions on the lattice,
it has generally been found that only relatively small gradient-flow amounts can be used, which in our context would be $\omega_n^2 \tflow \ll 1$.
This is discussed for instance in \cite{Eller:2018yje}.
Within this range the extrapolation towards zero flow is generally approximately linear (see \cite{Suzuki:2013gza} for a formal explanation of why this is expected), but if larger flow times are used it can become a complex and \textsl{a priori} unknown function.

However, we believe that the case of $\HEtf(\omega_n)$ is somewhat different.
After an amount of flow which from Figure \ref{fig:suscept} is $\tflow \lesssim 0.4a^2$, the total topology becomes fixed.
After this amount of flow, while $q(x,\tflow)$ remains flow-dependent, its integral $Q=\int \diff^4{x}\: q(x,\tflow)$ does not.
This is a special property of $q$ which arises because it is a topological density when evaluated on smooth configurations.
Therefore, further flow should be interpreted as a redistribution or smearing of $q(x)$, which in turn should result in a smearing of the correlation function 
\begin{equation}
    G(x) \to \int \diff^4{y}\: S(y) G(x-y) \,, \qquad
    \int \diff^4{y}\: S(y) = 1
\end{equation}
with $S(y)$ a unit-normalized smearing function, which in vacuum would be O(4) invariant and most likely nearly Gaussian.

The striking feature of such a smearing function is that, to the extent that $S(y)$ is O(4)-invariant, it has no effect on the correlation functions we are investigating.
Specifically, we can write \Eqref{HErotated} as
\begin{align}
    \HE(\omega_n) & = \int \diff^4{x} \: G(x) K_n(x) \,,    &
    K_n(x) = \frac{\cos(\omega_n \tau) \sinh(\omega_n r)}{\omega_n r} \,.
\end{align}
Applying a smearing to $G(x)$, this becomes
\begin{align}
    \HEtf(\omega_n) & \simeq \int \diff^4{x} \int \diff^4{y}
    \: G(x{-}y) S(y) K_n(x)
    = \int \diff^4{x} \: G(x) \int \diff^4{y} \: S(y) K_n(x{-}y)
\end{align}
where we used the symmetry of $S(y) \to S(-y)$ and integration-variable shifts to rewrite the expression in terms of a spatial averaging of the kernel $K_n(x)$.
But surprisingly, for any O(4)-invariant $S(y)$ function, our specific kernel has the property that
\begin{align}
    \int \diff^4{y} \: S(y) K_n(x-y) & = K_n(x) \,.
\end{align}
The easiest way to see this is to not perform the averaging over $\hat{k}$ in \Eqref{HErotated}, so 
$K_n(x) = e^{(z+i\tau) \omega_n}$.
Interpreting $(z+i\tau)$ as a value in the complex plane, $K_n(z,\tau)$ is an analytic (in fact, entire) function, and its average over any circle in the $(z,\tau)$ plane therefore equals the value at the center of the circle.
Therefore the value of $\HEtf(\omega_n)$ is only flow-dependent to the extent that the smearing function $S(y)$ is anisotropic between space and time.
Any anisotropy in $S(y)$ must arise due to the time-periodicity and must scale as $T^2$ or a higher power; therefore it is a $\tflow T^2$ effect.
Since the effect of flow is already suppressed by $\tflow$, this argument implies that the first flow effects in fact emerge at order $\tflow^2$.

\section{Results}
\label{sec:results}

\subsection{Numerical results}

\begin{figure}[tb]
    \centering
   \includegraphics[width=0.8\linewidth]{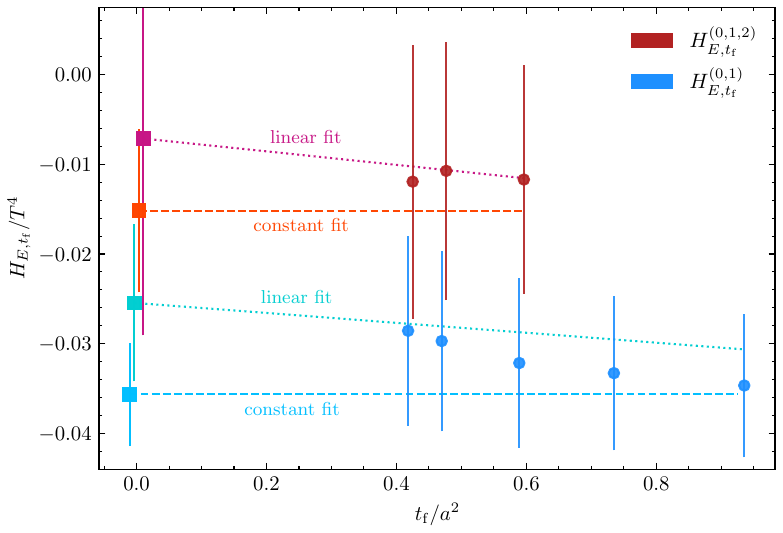}
    \caption{Plot of subtracted correlators $H_{E,t_\mathrm{f}}^{(0,1)}$ and $H_{E,t_\mathrm{f}}^{(0,1,2)}$ over flow time $t_\mathrm{f}$ (circular points), together with zero flow-time extrapolated values (square points) for constant and linear fits.}
    \label{fig:zero-tf-extrap}
\end{figure}

In the previous section we explained how we analyze lattice data to determine the correlation functions $\HE^{(0,1)}$ and $\HE^{(0,1,2)}$.
Here we will present our results.
We present our results at finite flow time in Figure~\ref{fig:zero-tf-extrap}, with the uncertainties obtained from 1000 bootstrap samples.
We argued in the previous section that the first flow-time dependence should arise at order $\tflow^2$, which for our chosen flow time range should be negligible.
Therefore it should be possible to fit the data over different flow times with a constant fit.
The figure shows such a fit for $\HE^{(0,1)}$ and for $\HE^{(0,1,2)}$.
Indeed, this fit works, with a good chisquared value.
What may come as a surprise is that the best fit actually lies below all of the data points.
This is a real effect caused by high correlations between different flow times, together with errors which grow at the smaller flow times.
We have checked that there are no relevant uncertainties introduced through determining and inverting the correlation matrix.
For completeness we have also included a fit where we allow for a linear-in-$\tflow$ coefficient.
This functional form also fits the data well, but the uncertainties are of course larger and the determined slope is consistent with zero within the fit errors.
Therefore the data does not support such a linear term, and we take the $\tflow$-independent fit to be our best final value.
We find
\begin{align}
    \HE^{(0,1)} & = -0.0357(58) \, T^4 ,
    &
    \HE^{(0,1,2)} & = -0.0152(91) \, T^4,
\end{align}
in SU(3) pure-glue gauge theory at $T=1.5 T_c$.
We emphasize that the negative sign is expected and arises because $q$ is an antihermitian operator; the relation to the production rate,
\Eqref{eq:nosubyet} carries a compensating minus sign.

Note that, in this exploratory study, we have only analyzed a single lattice spacing with $aT = 1/16$ ($N_\tau = 16$).
Formally there are lattice-spacing dependent, $\tflow$ dependent corrections, which could be of form $a^2/\tflow$ (though our use of an improved operator hopefully avoids this and leaves $a^4/\tflow^2$ corrections).
In a multiple-spacing study, such corrections would be eliminated by the continuum limit.
But because we know they are present, we could in principle have also carried out a $\tflow$ extrapolation including such corrections as well.
Since our data is well fit by a $\tflow$-independent constant, we presume that these corrections are under control in the current analysis.

\subsection{Comparison with analytical calculations}

Because we consider pure-glue QCD above $T_c$, it is not possible to compare with Ref.~\cite{DEramo:2021psx,Notari:2022ffe}, who consider pions in low-temperature full QCD.
However, we can compare with perturbative treatments, of which the most advanced is Ref.~\cite{Bouzoud:2026rur}.
They parametrize the susceptibility $\sigma(k,k)$ as:
\begin{equation}
    \sigma(k,k) = \frac{(N_c^2-1)\alphas^3(N_c+N_f/2) T}{3(4\pi)^2}
    \omega \: \gamma(k/T)
\label{NLOPT}
\end{equation}
where for us $N_c=3$ and $N_f$ is the number of fundamental quark flavors, which for us is $N_f=0$.
The numerical results and formulae for next-to-leading order perturbation theory in Appendix C of Ref.~\cite{Bouzoud:2026rur} show that $\gamma(k/T)$ varies weakly with $k$ in the range $k \sim 2\pi T$, the range most important in \Eqref{eq:sub-01} and \Eqref{eq:sub-012},
and its value in this range is around 8 to 9.
Taking $\gamma(k/T)$ to be constant in this range and adopting $\alphas \simeq 0.3$, we can substitute Eq.~(\ref{NLOPT}) into Eq.~(\ref{eq:sub-01}) and Eq.~(\ref{eq:sub-012}), with the result that
$\HE^{(0,1)} \sim - 0.036 \, T^4$ and $\HE^{(0,1,2)} \sim - 0.024 \, T^4$.
These are in fact consistent with our results.
Therefore we have verified that the next-to-leading order perturbative results found by Ref.~\cite{Bouzoud:2026rur} are consistent -- unlike the LO results, which are much too small to describe our findings.
The ratio $\HE^{(0,1,2)}/\HE^{(0,1)}$ must be less than 1 because the integral weight $h^{(0,1,2)}(\omega') < h^{(0,1)}(\omega')$.
We find a ratio somewhat smaller than the ratio of $2/3$ which arises if $\gamma(k/T)$ is constant, which can be interpreted as an indication that $\gamma(k/T)$ is an increasing function of $k/T$ in the range around $k=10T$ where the functions are the most different.
However, our error bars are too large to make a robust statement.

After the numerics of this study were completed, Ref.~\cite{Bonanno:2026ocw} became available, where the axion production efficiency is computed in pure-glue QCD on the lattice at a range of specific momenta via analytical continuation using the Backus-Gilbert method \cite{Backus:1968svk}.
Unfortunately their study used a different temperature than ours ($T=1.24 T_c$ rather than $T=1.5 T_c$).
It would be very interesting to use both approaches at the same temperature, to serve as a check on the validity of the Backus-Gilbert analytical continuation by seeing whether the rate determined by Backus-Gilbert continuation satisfies the integral constraints determined via our technique.

\section{Discussion}
\label{sec:discussion}

The spectral function determining axion production is given in \Eqref{eq:diffGamma}.
Our central result is that we can write a set of integral constraints on this axion production rate, shown in
\Eqref{eq:sub-01}, \Eqref{eq:sub-012} with $\HE(\omega_n)$ defined in \Eqref{HErotated},
and that the relevant Euclidean correlation functions can be successfully computed on the lattice.
As a test, we have presented such a computation in pure-glue QCD at one temperature and without a continuum extrapolation, showing that relatively modest numerical effort can determine one integral moment accurately and place useful constraints on the other.

One downside of our approach is that we fail to determine
$\diff{\Gamma}/\diff^3{k}$ directly as a function of $k$.
Instead, we only determine one or two integral moments of this quantity.
It is not clear to us how constraining this will be phenomenologically, or precisely how it should be best used in converting limits on the number of radiation degrees of freedom into limits on the axion mass.
This is certainly an interesting problem to address in the future.

Next let us discuss the prospects for repeating our measurement in full QCD at a range of temperatures around $T \in [200\:\text{MeV},1000\:\text{MeV}]$, where the level of equilibration of the axion most sensitively affects the final amount of cosmological radiation.
The observable we consider is bosonic and the analysis is relatively cheap, so it should be straightforward for anyone with a completed ensemble of finite-temperature full-QCD configurations to carry out a similar analysis.
Based on our experience here, there are two features (besides the availability of high statistics from a large ensemble) which are needed to make the analysis successful.
First, it was important that we worked on a rather fine lattice, with the number of lattice points in the time direction $N_\tau \geq 16$.
If this number is smaller, one may not have a range of gradient-flow depths which simultaneously satisfy $\tflow \geq 0.4 a^2$, needed for topology to be under control, and $\tflow m^2 \leq 1$, so the tail can be captured without flow artifacts.
Still-finer lattices might even allow the measurement of yet another integral moment.
Second, it was important that the correlator masses turned out to be significantly larger than the associated Matsubara frequencies:
in Table~\ref{masses-table-new}, $m_0/T > 0$, $m_1/T > 2\pi T$, and $m_2/T > 4\pi T$ each by a substantial amount.
This ensured that the decaying data outpaced the rising envelope function in \Eqref{sub-corr-indi}.
Without this feature, we would have to carry out the $r$-integral to a larger value, where the exponentially growing envelope would severely amplify the noise in the data.
Therefore the lattice determination is only feasible in situations where the lightest flavorless $0^{-+}$ state is sufficiently heavy compared to the temperature scale in the two or three lowest Matsubara sectors.
At weak coupling we expect $m_0$ to be of order $gT$ \cite{Arnold:1995bh}, which gets smaller as the temperature gets higher and $g$ gets smaller.
The differences $m_1-2\pi T$ and $m_2-4\pi T$ are presumably also parametrically of order $gT$, though we are not aware of a detailed calculation which supports this.
This implies that a lattice determination at very high temperatures may actually be more difficult than a determination closer to $T_c$, where the $\eta'$ is known to be heavy.

\section*{Acknowledgments}
We would like to thank Dietrich B\"odeker, who drew our attention to this problem, and Jonas Winter, with whom we had many fruitful conversations.
We also thank Killian Bouzoud, Jacopo Ghiglieri, and Eamonn Weitz for discussions after a presentation of a preliminary version of this work.
This research was funded by the DFG (Collaborative Research Center CRC-TR 211 ``Strong-interaction matter under
extreme conditions''~--~project number \mbox{315477589~--~TRR 211}).
Calculations were conducted on the Lichtenberg high performance computer of the TU Darmstadt.
This work was in part based on the MILC
collaboration's public lattice gauge theory code. \\
See {\hypersetup{urlcolor=black}\url{https://github.com/milc-qcd/milc_qcd}}.

\bibliographystyle{unsrt}
\bibliography{allecites.bib}

\end{document}